\documentclass[peerreview]{IEEEtran}
\usepackage{graphicx}
\usepackage{amsfonts}
\usepackage{amssymb}
\usepackage{amsmath}
\usepackage{newclude}
\usepackage{gastex}
\usepackage{url}
\usepackage{verbatim}
\usepackage{stfloats}
\usepackage{subcaption}
\usepackage{caption}
\usepackage{subfig}

\usepackage{subfig}
\usepackage{algorithm}
\usepackage{algpseudocode}
\usepackage[normalem]{ulem}

\DeclareGraphicsExtensions{.pdf,.jpeg,.png}
\graphicspath{ {plots/} }

\begin{document}
\title{On Measuring the Geographic Diversity of \\ Internet Routes}

\author{
  Attila~Csoma,
  Andr\'as~Guly\'as,
  L\'aszl\'o~Toka
  \thanks{A. Csoma is with Budapest University of Technology and Economics, Hungary and with HSNLab, Dept. of Telecommunications and Media Informatics. e-mail: csoma@tmit.bme.hu}
  \thanks{A. Guly\'as is with  Budapest University of Technology and Economics, Hungary, with HSNLab, Dept. of Telecommunications and Media Informatics and with MTA-BME Information Systems Research Group. e-mail: gulyas@tmit.bme.hu}
  \thanks{L. Toka is with  Budapest University of Technology and Economics, Hungary, with HSNLab, Dept. of Telecommunications and Media Informatics and with MTA-BME Information Systems Research Group. e-mail: toka@tmit.bme.hu}
  \thanks{The work of L\'aszl\'o Toka was supported by the J\'anos Bolyai Research Scholarship of the Hungarian Academy of Sciences (MTA).}
  }

\maketitle

\begin{abstract}

  Route diversity in networks is elemental for establishing reliable,
  high-capacity connections with appropriate security between endpoints. As for
  the Internet, route diversity has already been studied at both Autonomous
  System- and router-level topologies by means of graph theoretical disjoint
  paths. In this paper we complement these approaches by proposing a method for
  measuring the diversity of Internet paths in a geographical sense. By
  leveraging the recent developments in IP geolocation we show how to map the
  paths discovered by traceroute into geographically equivalent classes. This
  allows us to identify the geographical footprints of the major transmission
  paths between end-hosts, and building on our observations, we propose a
  quantitative measure for geographical diversity of Internet routes between any
  two hosts.
  
\end{abstract}

\begin{IEEEkeywords}
\\geodiversity; traceroute; geolocation; disjoint routes
\end{IEEEkeywords}

\section{Introduction}

The value of knowledge of the Internet topology is arguably immense. In the last
decades we have witnessed a myriad of stories in which topology-related
information about the Internet was directly compiled into more efficient
architectures, services and more appropriate business decisions. Content
Delivery Networks (CDNs) \cite{pathan2008taxonomy}, in which global topological
peculiarities are highly exploited for e.g. surrogate server and cache placement
strategies or request routing mechanism, are just a narrow segment of the whole
spectrum. Peer-to-peer networks \cite{castro2003topology,lua2005survey}, data
center placement \cite{greenberg2008cost}, traffic engineering
\cite{awduche2002overview}, business-based AS peering strategies
\cite{clark2011interconnection}, just to mention a few, are all receivers of
Internet topology related knowledge. With this non-comprehensive list of
receivers in mind, it should come at no surprise that many researchers have
contributed to our current understanding of the topology of the Internet.

An elemental metric of Internet topology is the diversity of routes between
end-hosts, as multiple uncorrelated routes can provide better throughput,
resiliency and security. In \cite{sterbenz2010resilience},
\cite{sterbenz2013evaluation} and \cite{rak2015resilient} authors describe how
to increase the resilience of future networks and the role of multipath
communication in it. A detailed description about network security can be found
in \cite{kaufman2002network} and authors in \cite{lou2003spread} propose a
method which improves network security; however it requires multiple path
between end-hosts.

Existing studies of IP-level route diversity usually focus on extracting routes
between end-hosts, e.g. by using \texttt{traceroute}, and on computing their
diversity by means of edge or node disjointedness in a graph theoretical sense
\cite{schwartz2010diversity,muhlbauer2006building, teixeira2003search}. Such
analysis provides an interpretation of route diversity in a microscopic level
where each node in the route is a router interface having a particular IP
address. In \cite{shavitt2012geographical} the authors propose to interpret
route diversity at a higher level, namely at the level of PoP's (Point of
Presence): the interfaces residing in the same building or campus are grouped
together, forming a PoP, and finally route diversity is computed at the level of
these PoP's. In this paper we interpret route diversity at an even higher,
geographical level. We propose grouping routers in a given geographic vicinity
and compute route diversity at the level of geographical regions (e.g. the level
of cities) independently from ASes. Our contribution is threefold: \textit{first
  we describe a method for identifying the geographically equivalent routes in
  \texttt{traceroute} outputs; second, we show the efficiency of the method in
  terms of successfully merged \texttt{traceroute} routes and present their
  possible applications; finally we define a metric which can capture the
  geo-diversity of Internet routes between endpoints and compute this metric for
  our measurement dataset}. Such characterization of routes' ``geo-diversity''
is clearly beneficial if one is curious about connectivity between end-hosts in
case of e.g. power outages affecting larger geographical areas.

The rest of the paper is organized as follows. In Sec.~\ref{related} we overview
the corresponding related work. Sec.~\ref{sec:measurement} describes our
\texttt{traceroute} measurements and our algorithms for extracting
geographically equivalent routes from those. In Sec.~\ref{sec:sim} we define and
evaluate a metric called Geographic Diversity Index (GDI) that captures how
Internet routes differ from each other. In Sec.~\ref{sec:results} we validate
our framework and present its performance. Finally we draw the conclusions and
list the possible applications and future work in Sec.~\ref{sec:conclusion}.
  
\section{Related work}\label{related}

Numerous existing studies apply geographical information to uncover non-trivial
aspects of the Internet topology. In \cite{subramanian2002geographic} the
authors use the geographical positions of routers to estimate route
circuitousness, route length distribution and geographic fault tolerance from an
end-to-end and from an ISP perspective. However, the DNS name-based geolocation
method used in their work has its own limitations and may create a false spatial
distribution as described in \cite{zhang2006dns}. The distance and angle between
consecutive IP hops are investigated in \cite{laki_detailed_2009}. In
\cite{boguna2010sustaining} the authors use geographic information to construct
the hyperbolic map of the Internet and prove the navigability of the AS level
topology using greedy algorithms. Points of presence are detected using delay
constraints on an IP interface graph and the distribution of PoPs around the
globe is visualized in \cite{shavitt2012geographical}. Authors of
\cite{shavitt2010structural} used PoP detection to evaluate the accuracy of some
IP geolocation database. Inter-AS route diversity is examined through the
network of Sprint in \cite{teixeira2003search}. In \cite{han2006experimental}
the authors study the route diversity of multi-homed and overlay networks as
seen from multiple vantage points using graph theoretical methods exclusively.
However none of these works capture route diversity in the pure geographical
sense on the router-level.

\section{Measurement framework}
\label{sec:measurement}

We built a framework that determines the extent of geographical heterogeneity of
end-to-end routes that are being used to carry traffic between any two points in
the Internet. The method that we implemented is the following: first, we run
\texttt{traceroute} measurements to collect the IP-level routes between the
selected endpoints; second, we use \textit{MaxMind} \cite{maxmind}, a
localization tool that determines the geographical position of the recorded IP
addresses; third, we group those IP-level routes that we consider equivalent
from a geographical perspective; finally, we calculate a route diversity index
for the selected endpoints. In this section we present theses steps in details.

\subsection{Route measurements with \texttt{traceroute}}

Usually \texttt{traceroute} is used to discover end-to-end routes between two
endpoints in the Internet. Network operators use it for detecting network
errors, researchers use it to build Internet topology models. Although most
in-network routers and endpoints support its operation, \texttt{traceroute} has
a number of well-known shortcomings. On one hand it can be easily deceived by
load-balancers, on the other hand it is an active measurement tool and due to
the extra data traffic it generates, certain network equipment are configured to
disable reactions to \texttt{traceroute} (and \texttt{ping}).

Several projects exist that collect \texttt{traceroute} measurements and make
them publicly available. These data sources differ significantly based on their
vantage point types, their vantage point location, the \texttt{traceroute}
implementation they run and their endpoint selection methods. Two such projects
are IPlane \cite{iplane} and CAIDA's ITDK (Internet Topology Data Kit)
\cite{caida_itdk}. IPlane offers a route performance prediction service and
periodically runs \texttt{traceroute} measurements from PlanetLab
\cite{Chun_CCR_2003} nodes to a set of endpoints changed bi-weekly. ITDK data
sets are produced from measurements collected by CAIDA's Archipelago project:
\texttt{paris-traceroute} \cite{augustin_avoiding_2006} is run from 89 vantage
points spread over 37 countries to randomly selected endpoints. In order to
measure route diversity between two endpoints, we need to detect as many routes
between those two endpoints as possible. Although the \texttt{paris-traceroute}
output of ITDK is more reliable than that of IPlane's \texttt{traceroute}, the
random selection of endpoints implemented by CAIDA hinders the collection of
routes between the same vantage- and endpoints. Therefore we used the data of
IPlane's \texttt{traceroute} measurements.

\subsection{IP localization and filtering of routes}
\label{sec:filtering}

Once we have the IP-level routes, we determine the geographic position of each
node appearing in them. Naturally the accuracy of the positioning is of
paramount importance. As it is also noted in \cite{laki_spotter:_2011,
  zhang2006dns}, the use of DNS names and contents of various registries leads
to unacceptable inaccuracy. Instead, we use the freely available geolocation
tool MaxMind GeoLite \cite{maxmind} in order to establish the position for the
IP addresses recorded in the measurements. As pointed out in \cite{geocompare},
it is one of the most reliable, freely available geolocation database.

We filter the \texttt{traceroute} dataset from IPlane in order to remove
measurements of vantage- and endpoint pairs between which only 1 IP-level route
was observed. After localizing the IP hops, we further removed those vantage-
and endpoint pairs between which only 1 geographic path was observed. With this
second filter we eliminated traces differing only due to IP-aliasing and load
balancing. The remaining set of traces contained $\sim0.5$ million discovered
routes; interestingly $\sim80$\% of vantage- and endpoint pairs had only 1
geographic path in the measurements.

\subsection{Clustering of routes}
\label{sec:geodiverse-routes}

After establishing \textit{geo-paths} for the remaining routes by localizing
their IP hops, we set out to decide which routes can be considered to be the
same and which ones are different in a geographical sense. We make this decision
by clustering the geo-paths on a hop-by-hop basis and by defining geodiversity,
a mutual metric, between geo-paths. Iterating through all the geo-paths from a
given vantage point to a given endpoint, we choose an appropriate cluster for
each geo-path: if the geo-path satisfies the geographical equality with all
cluster members then it is assigned to that cluster if not, then a new cluster
is created for it. At the end, all geo-paths in each cluster are considered to
be geographically the same.

Two geo-paths are geographically equal if they are not farther from each other
than 50 km: the distance between any of their nodes and the closest one of the
straight lines determined by consecutive nodes of the other geo-path is not
larger than 50 km. We used the arbitrary threshold of 50 km to reflect a large
city's diameter \cite{geoloc_db}. Note, that this threshold also allows for
fiber duct curves in the physical network, alleviating the mismatch between the
location of IP-level nodes of \texttt{traceroute} measurements and the actual
trail of the underlying physical links.

\begin{figure}[h]
	\centering
	\includegraphics[width=.4\linewidth]{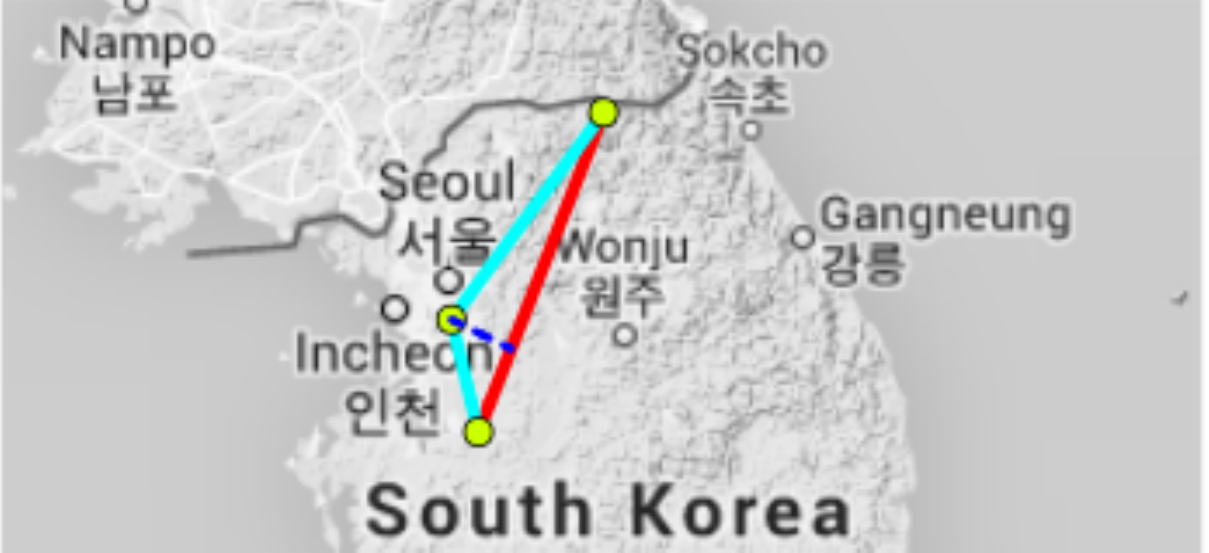}
	\caption{Route merging example}
	\label{fig:route_merging}
\end{figure}

As a demonstrative example, we drew two geo-paths that are grouped in the same
cluster in Fig. \ref{fig:route_merging}. Those routes differ at the IP-level and
their geo-paths are also different, but since the distance (marked with dashed
blue line) of the intermediate node of one route from the geo-path of the other
route is smaller than 50 km, our clustering algorithm rules the geographic
differSence between these two routes negligible.

\section{Calculation of geographic route diversity} \label{sec:sim} 

Our final step to capture the geographic route diversity is to define and
evaluate a Geographic Diversity Index (GDI). We require GDI to produce values to
a given route set between a source and destination pair such that multiple
geo-paths spanning over large geographical areas get a higher GDI.

\subsection{Requirements}\label{sec:req}
\begin{figure}[h]
  \centering
  \includegraphics[width=.32\linewidth]{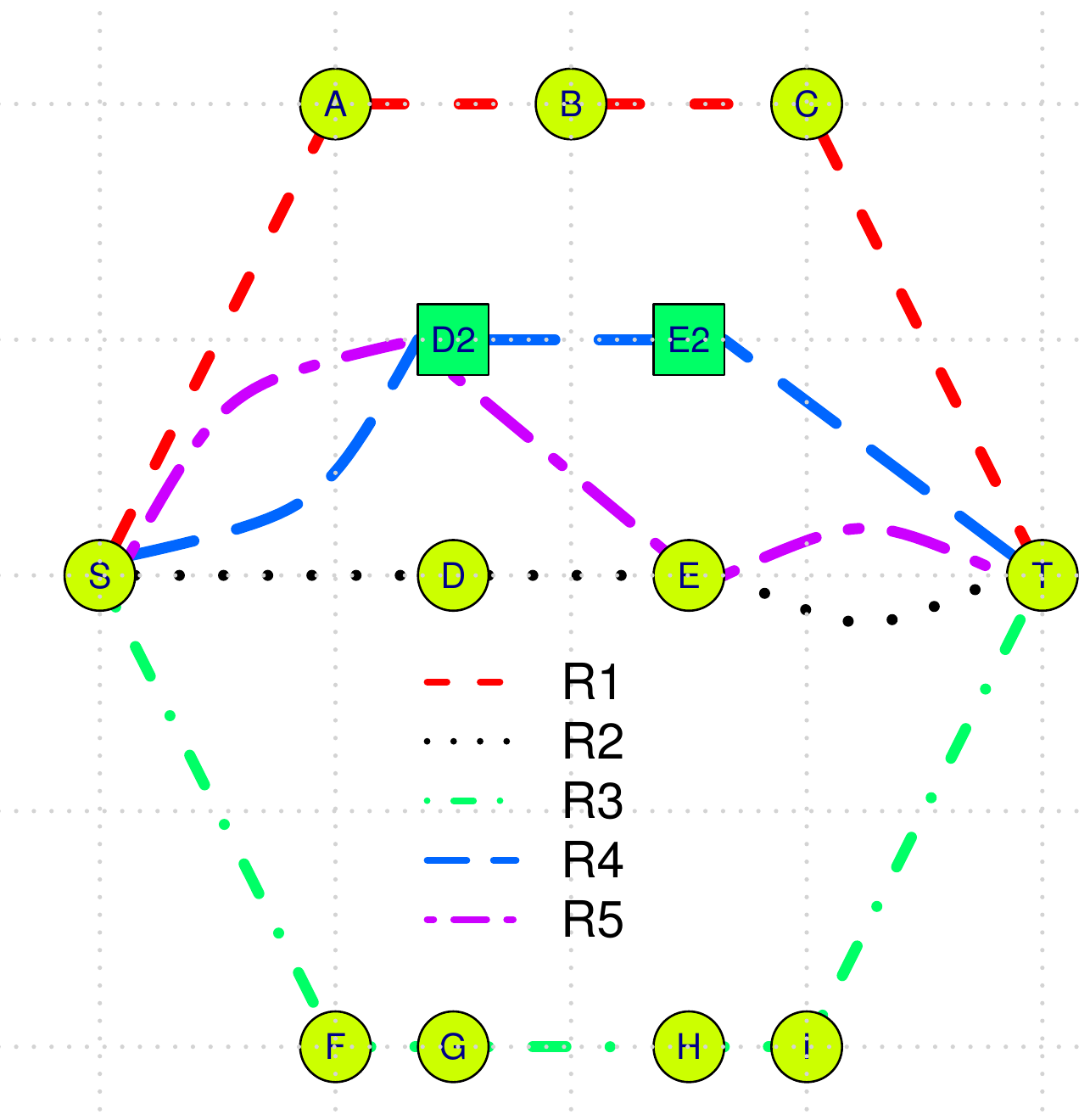}
  \caption{Geo-diversity example}
  \label{fig:route_example}
\end{figure}

Before describing the computation of GDI in details we highlight the key
attributes which make a route set ''more'' diverse in a geographical sense
against an other. Let us assume a source (S) and a target (T) node as endpoints
and two ''cover'' routes S-A-B-C-T (R1) and S-F-G-H-I-T (R3) as shown in
Fig.~\ref{fig:route_example}\footnote{Lines are curved to distinguish different
  routes using the same link.}. Let us assume that nodes are positioned
according to their geographic positions. Let us also assume that the GDI for
this setting is $\hat{r}$. Our first goal is to reward higher route count.
Therefore a route set with two routes must achieve lower GDI than the same route
set extended with another arbitrary route. That is if we add R2 to our
theoretical example and obtain $\hat{r}_2$ as the GDI for this amended route
set, then we expect $\hat{r}<\hat{r}_2$. Our second goal is to reward higher
geographic distance between routes. Therefore if we modify the route set by
reducing the distance between the routes, then the GDI of the new route set must
be lower: by replacing R2 with R4 (new GDI is $\hat{r}_4$), we require the GDI
to fall. It follows that geo-paths with varying distances (from other routes)
increase more the GDI (e.g. R5 and $\hat{r}_5$) than parallel geo-paths (e.g. R4
and $\hat{r}_4$) closer than R2. Therefore we require the order between GDI
values as:
\begin{equation}\label{eq:order}
  \hat{r}<\hat{r}_4<\hat{r}_5<\hat{r}_2
\end{equation}

\subsection{Geographic Diversity Index (GDI)}

\begin{figure}[b]
  \centering
  \includegraphics[width=.35\linewidth]{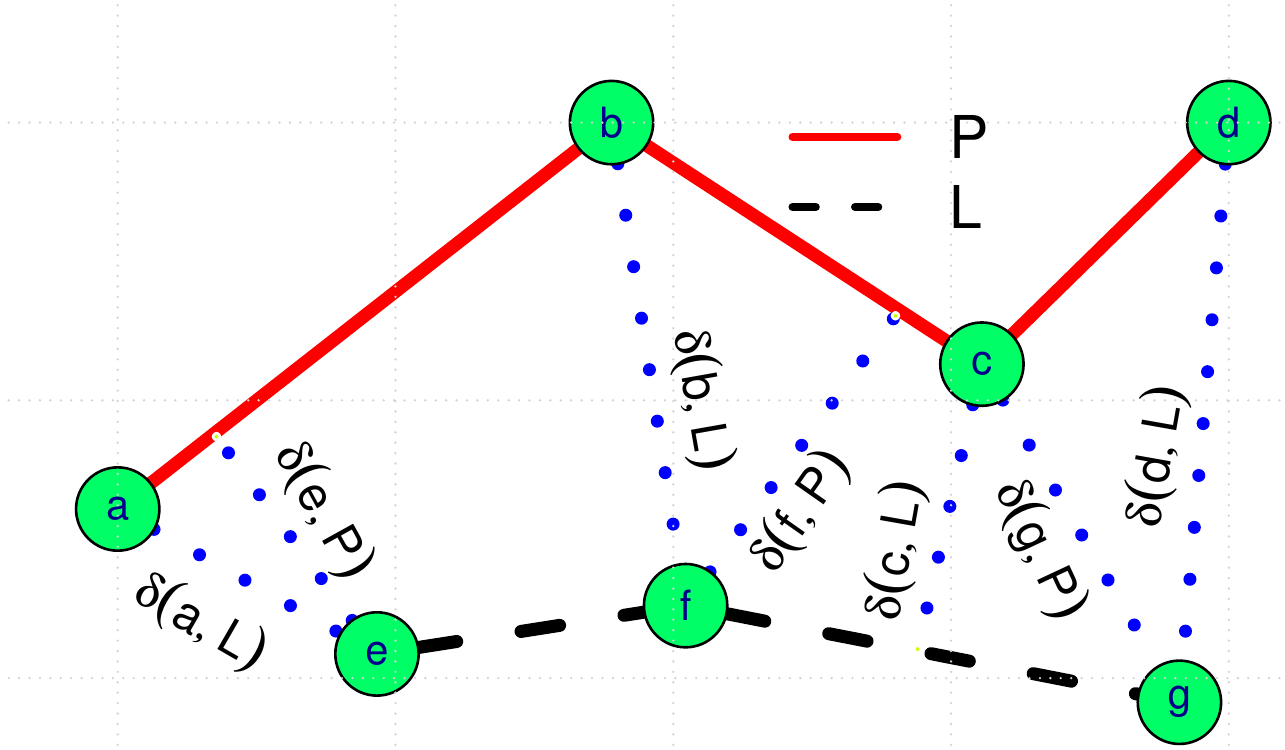}
  \caption{Distance example}
  \label{fig:distance_example}
\end{figure}

We model geo-paths of Internet routes as collections of sections between their
consecutive nodes. Let us assume two routes defined by their nodes: $P=\{a, b,
c, d\}$ and $L=\{e, f, g\}$. Naturally, the distance $ \delta(a,L)$ between node
$a$ and route $L$ translates to the distance between $a$ and the closest point
(not necessarily a node) of route $L$ to $a$. Let $\Delta (P, L)=\{\forall u \in
P| \delta(u,L), \forall u \in L| \delta(u,P)\}$ be a vector containing all
possible node distances between $P$ and $L$. In the toy example of
Fig.~\ref{fig:distance_example} $\Delta (P, L)=\{\delta(a,L), \delta(b,L),
\delta(c,L), \delta(d,L), \delta(e,P), \delta(f,P), \delta(g,P)\}$.
\begin{comment}
Naturally, the distance between node $s$ and route P translates to the distance
between $s$ and $t_s$ which is the closest point in route P to $s$. The distance
between route P and route L is defined by the sum of all their node distances
and it is denoted as $d(P, L)$.
\end{comment}

In order to satisfy the requirements listed in Sec. \ref{sec:req} we define GDI
for a given set of routes as follows. First, we define a diversity index between
two routes as:
\begin{equation}\label{eq:simi}
  \mathrm{d}(P, L) = \Big(1-\text{Var'}\big(\Delta(P, L)\big)\Big)
  \text{Mean}(\Delta(P,L)),
\end{equation}
where Var' denotes the variance of its arguments normalized to the interval
$[0..1]$. Second, we define the diversity between a single route $P$ and a set
of routes $\mathbb{V}$ as:
\begin{equation}\label{eq:dev}
\mathcal{D}(P,\mathbb{V}) = \min_{L \in \mathbb{V}}\mathrm{d}(P, L)
\end{equation}
Finally, we quantify the overall GDI for a given set of routes. In order to
calculate this, we use a step-by-step method. Let us assume a set $\mathbb{V}$
containing the routes. The process starts from an empty set $\mathbb{U}=\{\}$.
In the $0$. step we search in set $\mathbb{V}$ for the two routes having the
highest diversity score $\mathrm{d}_{0}=\max_{P,L \in \mathbb{V}}(\mathrm{d}(P,
L))$ and move paths $\text{argmax}_{P,L \in \mathbb{V}}(\mathrm{d}(P, L))$ to
$\mathbb{U}$. In the $i$-th step we compute $\mathcal{D}_{i}=\max_{P \in
  \mathbb{V}}(\mathcal{D}(P, \mathbb{U}))$ and move the path $\text{argmax}_{P
  \in \mathbb{V}}(\mathcal{D}(P, \mathbb{U}))$ to $\mathbb{U}$. The process
terminates when $\mathbb{V}$ is empty. Finally we compute GDI as:
\begin{equation}\label{eq:GDI}
\text{GDI} = \mathrm{d}_{0} + \sum \mathcal{D}_{i}.
\end{equation}

To demonstrate that the proposed method of GDI calculation satisfies the
requirements we set, we show the GDIs of the route sets defined in
Sec.~\ref{sec:req} (with a grid cell size is $\sim84$ km in
Fig.~\ref{fig:route_example}) in Table~\ref{table:gdi}. Note, that the produced
GDI values fulfill Eq. \ref{eq:order}.

\begin{center}
  \begin{tabular}{ l | r }
    $r$ & 2182 \\ \hline
    $\hat{r}_4$ & 3299 \\ \hline
    $\hat{r}_5$& 3467  \\ \hline
    $\hat{r}_2$& 3729 \\ 
  \end{tabular}
  
  \captionof{table}{GDI values for the geo-diversity example}
  \label{table:gdi}
\end{center}

\section{Results}
\label{sec:results}

In this section we show the geodiversity results we achieved from the
measurement data set. First we show through an illustrative example the
difference between the raw \texttt{traceroute} outputs and the geographically
equivalent routes achievable after applying our route clustering algorithm.
Second, we present the compression rates that we were able to attain on the
whole measurement set. Finally, we show the GDI results that we calculated for
the already clustered route set.

\subsection{An example for route clustering}

\begin{figure*}
\centering
\begin{subfigure}[t]{.49\columnwidth}
\includegraphics[height=.4\linewidth]{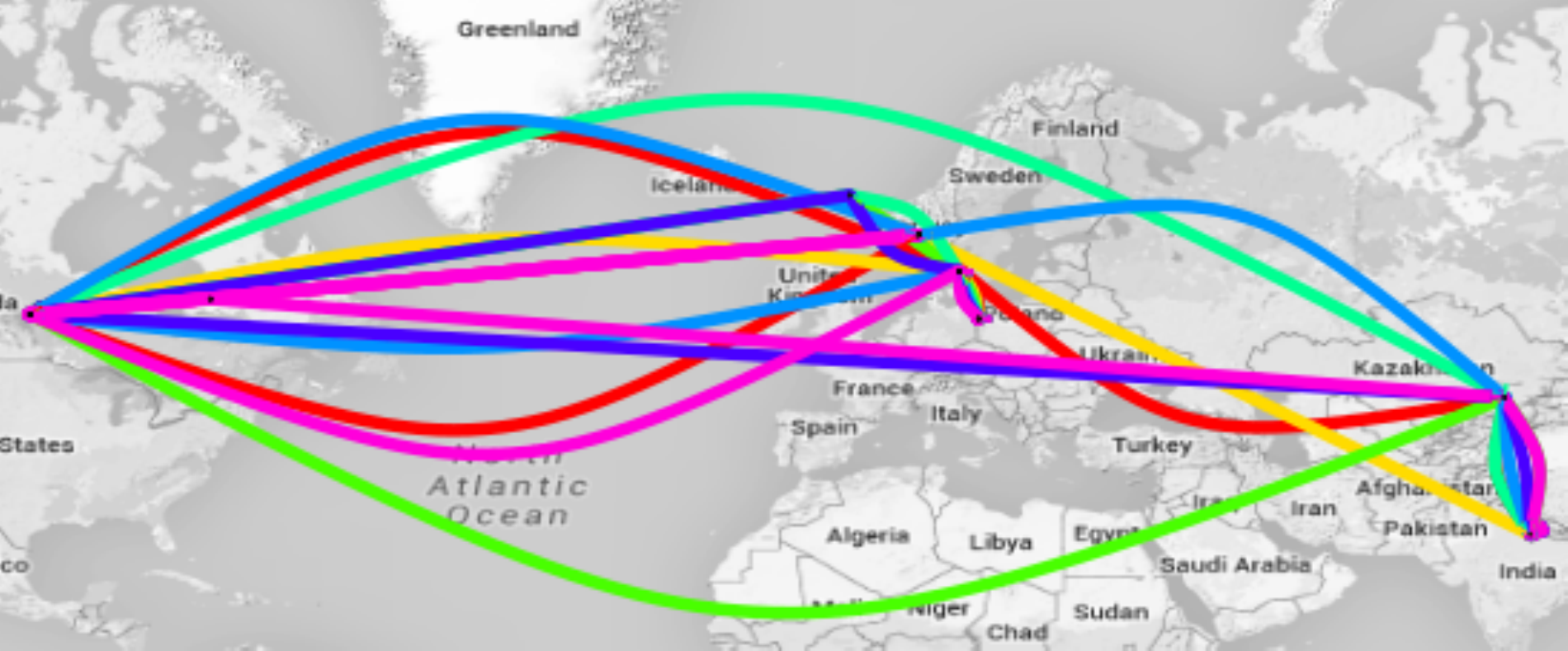}
\caption{\texttt{traceroute} output, route count: 7}
\label{fig:before-merge}
\end{subfigure}
\begin{subfigure}[t]{.49\columnwidth}
\includegraphics[height=.4\linewidth]{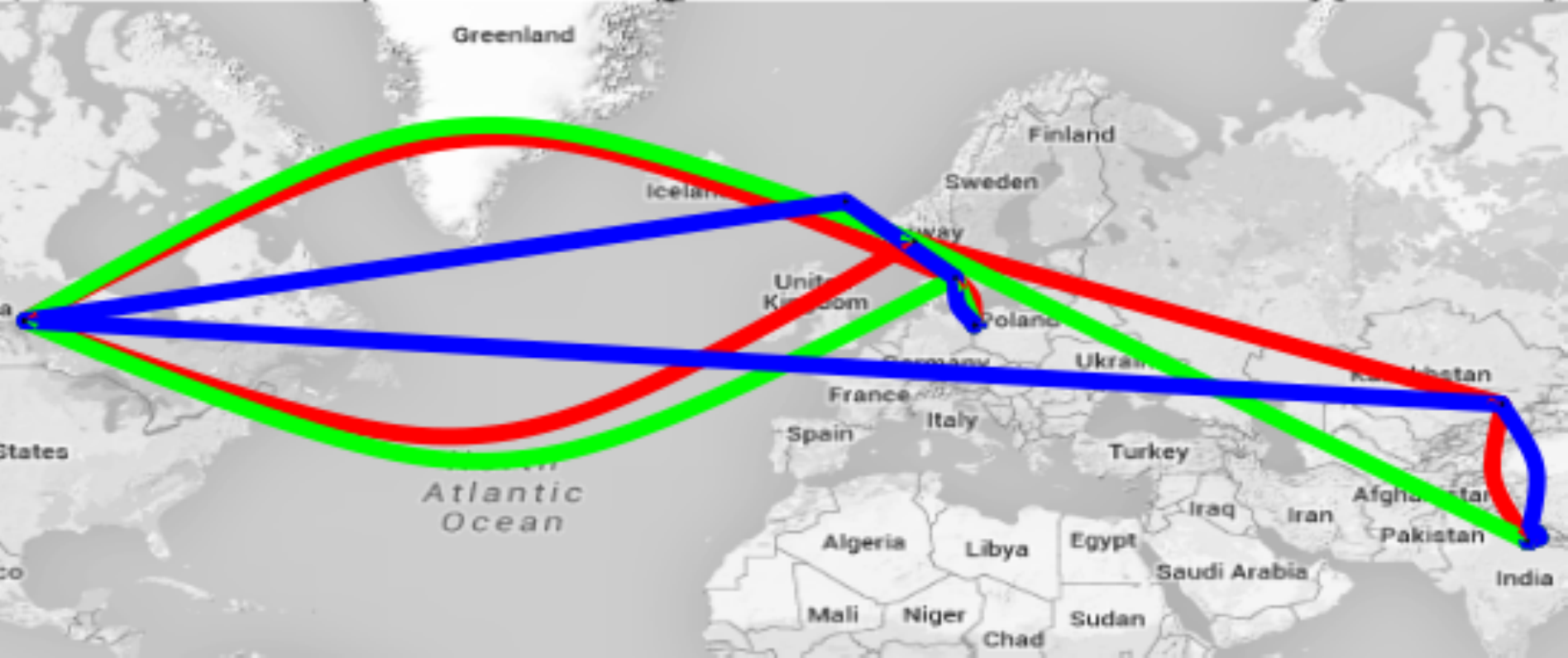}
\caption{Geodiverse routes, route count: 3}%
\label{fig:after-merge}
\end{subfigure}
\caption{Route count comparison}
\label{fig:route_count_example}
\end{figure*}

An example of the results of our route clustering algorithm is shown in Fig.
\ref{fig:route_count_example}. Arcs represent hops between the localized IP
nodes obtained from \texttt{traceroute} output. Note, that arcs do not indicate
a real link trajectories, merely distinguish routes on the same intermediate
links. On the left-hand side one can observe all the routes that the
\texttt{traceroute} measurements yielded between a source node located in Poland
and a destination node located in India (Fig.~\ref{fig:before-merge}). Between
these two hosts there exist 7 different routes on the IP-level, but only 3
geographically different routes, i.e., geo-paths (Fig. \ref{fig:after-merge}).

\subsection{Compression ratio of route counts}

\begin{figure}[h]
	\centering
	\includegraphics[width=.32\linewidth]{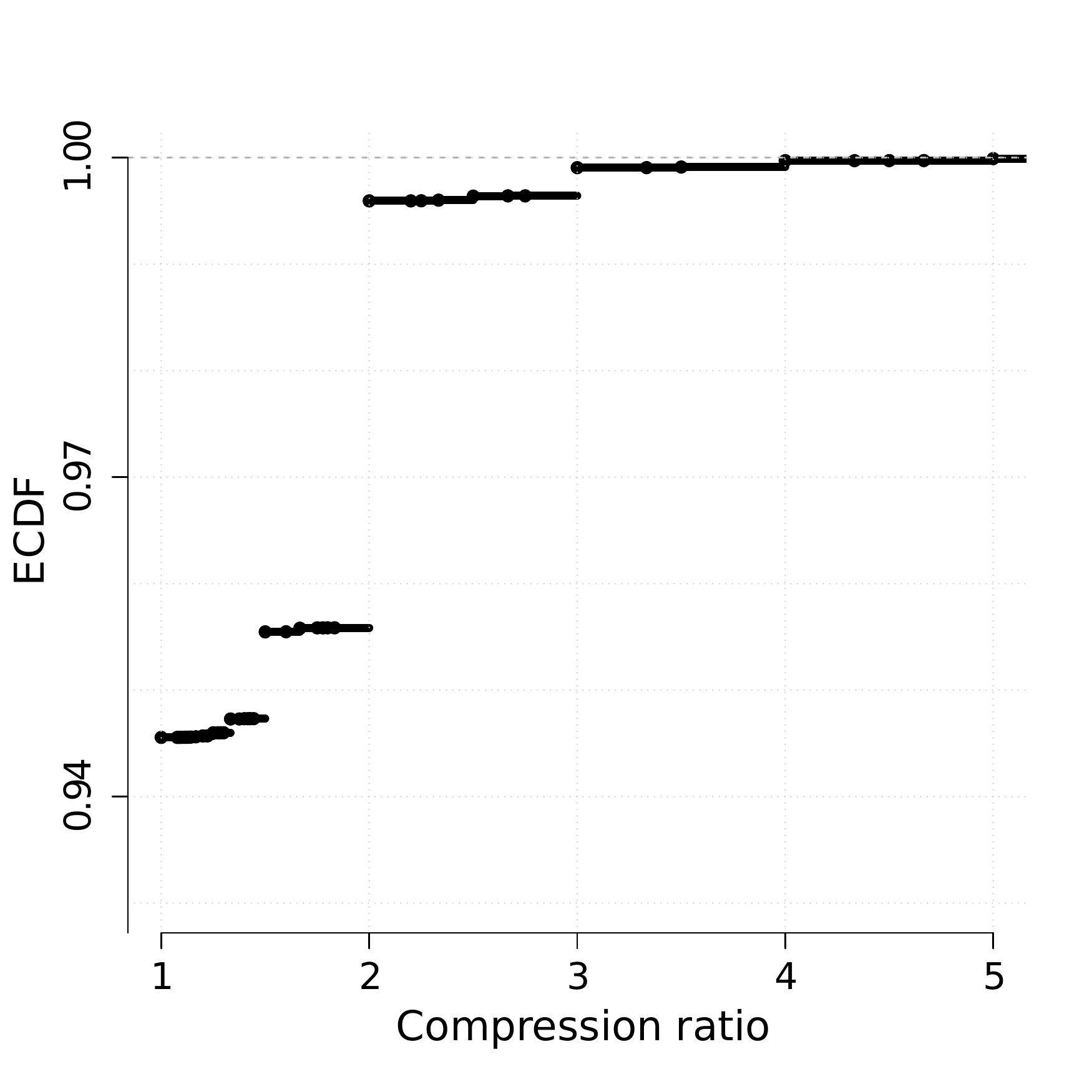}
	\caption{ECDF of compression ratios for all source-destination pairs}
	\label{fig:compression_all}
\end{figure}

Stepping up from one example, here we show the overall results in terms of route
clustering on the whole measurement data set. We call the ratio of the number of
original routes and the resulting clusters as the geo-compression ratio. In
Fig.~\ref{fig:compression_all} we plot the empirical distribution of this
geo-compression ratio for all source and destination host pairs in our data set.
  
The fraction of source-destination pairs for which the route clustering does not
decrease the number of geographically equivalent routes is large (around 80\%).
In these cases the main reason for the poor compression performance is the fact
that only one route is yielded by \texttt{traceroute}. Another reason, in less
significant number of cases, is that the multiple recorded routes run through
nodes at exactly the same locations (Sec. \ref{sec:filtering}). On the remaining
20\% of source-destination pairs we applied our clustering algorithm with
remarkable results: this is shown in Fig. \ref{fig:compression_all}. In more
than 4\% of the all cases we could obtain a geo-compression ratio higher than 2.

\subsection{Geo-diversity results}

\begin{figure}[]
	\centering
	\includegraphics[height=.32\linewidth]{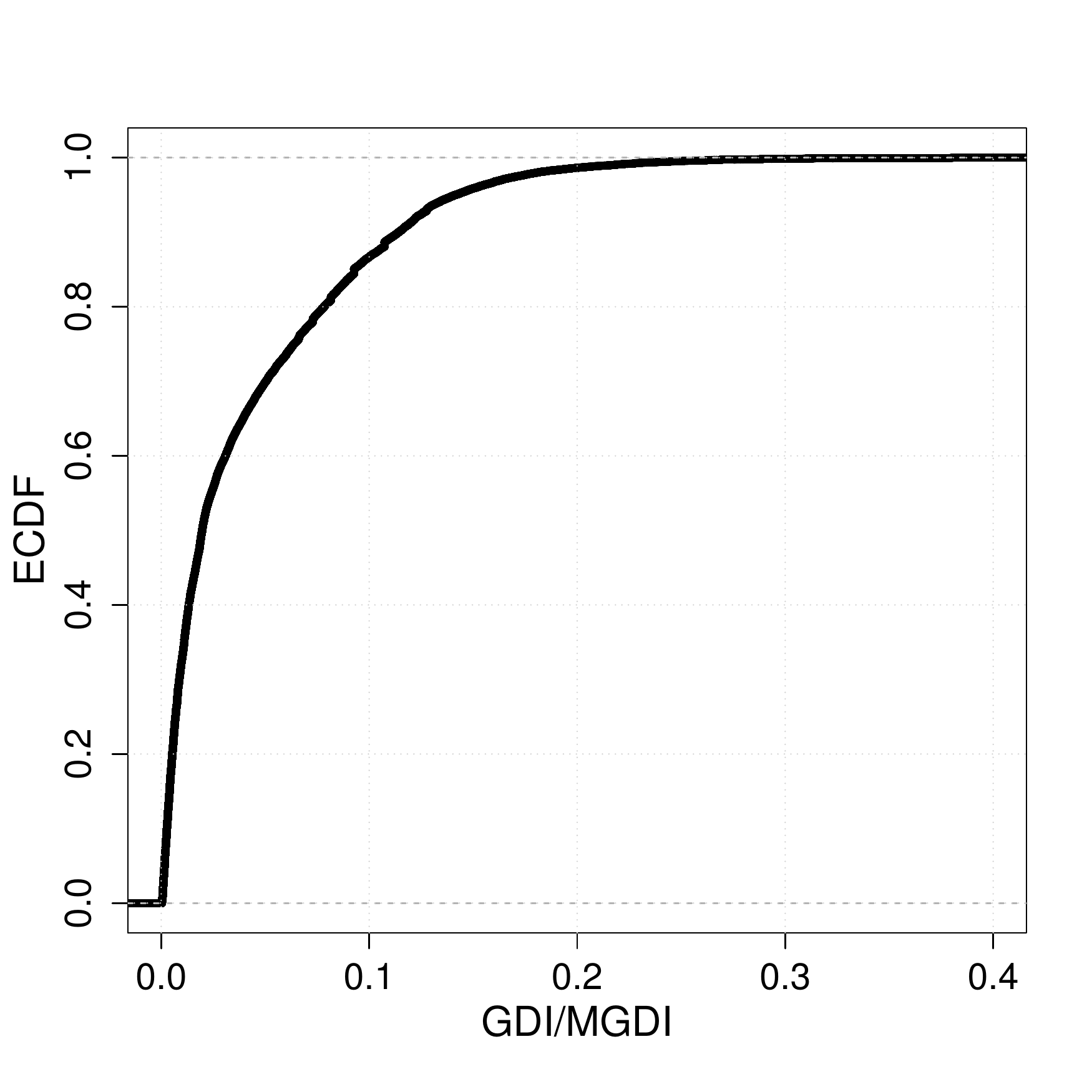}
	\caption{Geo-diversity results given by GDI/MGDI}
	\label{fig:prod_ref}
\end{figure}

Using the GDI metric that we defined in Sec. \ref{sec:sim} to characterize the
geo-diversity of routes between a source and a destination pairs we show how the
calculated values compare to the theoretical maximum of the same metric, taking
only the number of geographically equivalent routes and the length of the
longest one into account (not their actual trajectories). For this hypothetical
maximal value, denoted as MGDI, we place a number of routes forming triangle
shapes, the longest one reaching to the highest, so that their GDI would be the
largest. In Fig. \ref{fig:prod_ref} we plot the distribution of the ratios of
GDI over MGDI for those source-destination pairs between which we found at least
2 geographically different routes (20\% of all pairs, as mentioned above). The
results show that for 80\% of these cases the GDI of routes is less than 10\% of
their MGDI, i.e., the theoretically maximum diversity given the number of routes
and the length of the longest route.

\section{Conclusion}
\label{sec:conclusion}

Our goal was to find out where on Earth the packets travel exactly when traffic
is carried over the Internet. We discovered that between two given points packet
flows are not so scattered as the diversity of \textit{traceroute} results
suggest. We actually saw very few host pairs between which more than 1
geographically different routes exist. We showed that 80\% of endpoints with
more than one route has less than a 10\% MGDI value which indicates low
diversity in terms of geographic distance. The knowledge we gained from this
study about the geographical diversity of Internet routes is useful for several
applications. In this section we give a few examples of these, and we discuss
the weaknesses of our method.

\subsection{Applications}
\label{sec:applications}
\textbf{Estimating bufferbloat:} In order to measure Internet delay correctly
one must fight many sources of inaccuracy: if one-way latency measurement is
possible between two hosts, their clocks must be synchronized, if not, several
issues come up: misleading \textit{traceroute} results due to load balancing and
MPLS tunnels, different return path of the ICMP\_REPLY when using \textit{ping},
etc. Indeed, even if the topology is correctly discovered, many aspects of the
actual operation of the network equipment pieces affect the measured delay. In
order to somehow infer the impact of bufferbloat from the total delay, a very
hot topic nowadays, it is important to have an idea about the propagation delay
of the packets. Since the propagation delay is closely related to the traveled
geographic distance, the sets of geographically equivalent paths, discovered by
our method, provide important input to the analysis of the bufferbloat
phenomenon.

\textbf{Network resilience:} Network resilience, in its classical sense, is a
well-studied research domain \cite{najjar1990network}. When network links are
not going down individually, but instead are affected en masse due to a regional
catastrophe, let it be a natural disaster, a power blackout or an EMP attack,
then the geographic topology of the Internet suffers correlated link failures.
In order to be ready for this, planning geographic redundancy of Internet paths
can use the results of our method as an input.

\subsection{Discussion} 

After discussing the potential role of our method in various use cases, we turn
to the weaknesses of it that we are aware of. First, by applying measurements
created by the relatively simple tool \textit{traceroute}, we do not tackle
IP-aliasing when building the paths before clustering nodes close-by to each
other. One could argue that performing an already documented merging method
targeting IP aliases might yield the same result as the geomerging we do.
Second, one might question the accuracy of MaxMind, the tool we use to position
the nodes. However its accuracy is explored in details in \cite{geocompare} and
in long run measurements, we plan to use active measurement based geolocation
tools, such as Spotter \cite{matray_spatial_2012}. Third, it can be argued that
estimating geo-paths using straight lines between IP-level nodes may be
misleading. However, our city-sized threshold ensures that as long as there are
IP-level nodes in close proximity of fiber ducts' ends, this is avoided with
high probability. Finally, the proposed GDI and MGDI metrics might seem
simplistic, but we argue that to define a diversity metric between routes, many
key attributes must be taken into account, and a trade-off between various
features and scenarios must be accepted.

\bibliographystyle{IEEEtran}
\bibliography{ref}

\end{document}